\documentclass[
 aps,
 prb,
 preprint,
 superscriptaddress,
 amsmath,amssymb,
 longbibliography
]{revtex4-2}

\usepackage{hyperref}

\usepackage{graphicx}
\usepackage{dcolumn}
\usepackage{color}
\usepackage{hyperref}

\begin{document}

\title{Comparative density functional theory study for predicting oxygen reduction activity of single-atom catalyst}

\author{Azim Fitri Zainul Abidin}
\affiliation{
Department of Precision Engineering,
Graduate School of Engineering,
Osaka University,
2-1 Yamadaoka, Suita, Osaka 565-0871, Japan
}

\author{Ikutaro Hamada}
\email{corresponding author: ihamada@prec.eng.osaka-u.ac.jp}
\affiliation{
Department of Precision Engineering,
Graduate School of Engineering,
Osaka University,
2-1 Yamadaoka, Suita, Osaka 565-0871, Japan
}

\date{\today}

\begin{abstract}
It has been well established that nitrogen coordinated transition metal, TM-N$_{4}$-C (TM$=$Fe and Co) moieties, are responsible for the higher catalytic activity for the electrochemical oxygen reduction reaction.
However, the results obtained using density functional theory calculations vary from one to another, which can lead to controversy.
Herein, we assess the accuracy of the theoretical approach using different class of exchange-correlation functionals, i.e., Perdew-Burke-Ernzerhof (PBE) and revised PBE (RPBE), those with the Grimme's semiempirical dispersion correction (PBE+D3 and RPBE+D3), and the Bayesian error estimate functional with the nonlocal correlation (BEEF-vdW) on the reaction energies of oxygen reduction reaction on TM-N$_{4}$ moieties in graphene and those with OH-termination.
We found that the predicted overpotentials using RPBE+D3 are comparable and consistent with those using BEEF-vdW.
Our finding indicates that a proper choice of the exchange-correlation functional is crucial to a precise description of the catalytic activity of this system.
\end{abstract}

\maketitle

\clearpage

\section{Introduction}
\label{sec:introduction}
Single-atom catalysts are an important type of the catalysis that have gain considerable attention for the electro-reduction of $\mathrm{O_{2}}$ due to their potential to reduce the cost of electrochemical energy conversion devices such as fuel cells and metal-air batteries \cite{lefevre2009iron,lin2014noble,xu2018universal,wang2019achievements}.
Among explored catalyst formulations, non-precious metal, such as Fe and Co, embedded in N-doped graphene (Fe-N-C and Co-N-C) is the most representative of single-atom catalysts that have been shown to have remarkable catalytic activity and selectivity against oxygen reduction reaction (ORR) \cite{wu2011high,chen2017inside,xie2020performance,fitri2017cobalt,abidin2018effect,abidin2019nitrogen,sudarsono2022elucidating}.
Recent experimental results report that the highest ORR catalytic activity of the Fe-N$_{}$-C and Co-N$_{}$-C catalysts measured from the half-wave potentials ($E_{\mathrm{1/2}}$) is $\sim$ 0.83 V and $\sim$ 0.75 V, respectively \cite{yin2020construction,xiao2019climbing}.
This class of materials exhibits unique structures and properties that can potentially match and compete with the performance of platinum.
%
The mechanism of ORR on these catalysts has been extensively investigated: Among different N/C environments surrounding the metal atom, a four-fold coordinated metal atom with pyridinic nitrogen atoms in graphene (such Fe-$\mathrm{N_{4}}$-C and Co-$\mathrm{N_{4}}$-C moieties) is proposed to be the active site, due to their optimal binding strength with the chemical species involved in the ORR process \cite{zitolo2015identification,zitolo2017identification}.
The reaction energies and the intermediate reactions on the Fe-$\mathrm{N_{4}}$-C and Co-$\mathrm{N_{4}}$-C sites have been extensively studied using periodic density functional theory (DFT) calculations based on the computational hydrogen electrode (CHE) model \cite{kattel2013catalytic,kattel2014density}.
%
%
The overpotential for ORR ($\eta_{\mathrm{ORR}}$) estimated from theoretical limiting-potential in the CHE model is one of the most useful metric for screening the catalysts for the ORR \cite{norskov2004origin,kulkarni2018understanding}.
However, despite of the high ORR catalytic activities of the Fe-$\mathrm{N_{4}}$-C and Co-$\mathrm{N_{4}}$-C catalysts observed in the experiments, the $\eta_{\mathrm{ORR}}$ predicted through the CHE model are scattered.
Kattel \textit{et al.}\cite{kattel2014density} and Li \textit{et al.}\cite{li2015atomic} estimated the $\eta_{\mathrm{ORR}}$ values of 0.91 V and 0.23 V for Fe-$\mathrm{N_{4}}$-C and Co-$\mathrm{N_{4}}$-C, respectively, using the Perdew-Burke-Ernzherof (PBE) generalized gradient approximation (GGA) functional.
Sun \textit{et al.} estimated the $\eta_{\mathrm{ORR}}$ values of 0.79 V and 0.29 V for Fe-$\mathrm{N_{4}}$-C and Co-$\mathrm{N_{4}}$-C catalysts,  respectively, using PBE with the Grimme's semiempirical dispersion correction (DFT-D2)\cite{sun2019itinerant}.
Wang \textit{et al.} included the solvation effect through continuum solvation model with PBE and DFT-D, and obtained the $\eta_{\mathrm{ORR}}$ values of the 0.72 V and 0.83 V for Fe-$\mathrm{N_{4}}$-C and Co-$\mathrm{N_{4}}$-C, respectively \cite{wang2020axial}.
In addition to the solvation effect, they also included the effect of OH-termination as suggested by Wang \textit{et al.} \cite{wang2019self}, resulting in lower $\eta_{\mathrm{ORR}}$'s of 0.59 V and 0.74 V for Fe-$\mathrm{N_{4}}$-C and Co-$\mathrm{N_{4}}$-C, respectively.
However, a previous paper by the same authors reported $\eta_{\mathrm{ORR}}$'s of 0.47 V and 0.75 V for the Fe-$\mathrm{N_{4}}$-C and Co-$\mathrm{N_{4}}$-C through implicit solvation model with PBE, respectively \cite{wang2019self}.
%
%
Apparently, there is a discrepancy among the DFT studies besides the inclusion of dispersion correction as well as the solvation effect.
%
%

%
Theoretical studies based on the DFT method have been conducted using various approaches and have proven useful to elucidate the mechanisms of the reactions, especially in heterogeneous catalysis systems.
However, the accuracy and thus the success of DFT-based studies depend on the choice of exchange-correlation (XC) functional \cite{kirchhoff2021assessment}.
However, GGA functionals are prone to self-interaction errors for non-metallic system and calculation of accurate binding energies of the adsorbates on transition metal surfaces are a long-lasting challenge for the DFT XC functional \cite{sham1983density,perdew1985density,perdew1983physical,seidl1996generalized,patra2019rethinking}.
Improper choice of the XC functional can result in inaccurate binding energies and thus, the incorrect reaction mechanism.
Moreover, it is known that the van der Waals (vdW) or dispersion interactions, which can be important in the catalytic reactions, cannot be captured properly by GGAs \cite{hamada2012adsorption}.

In this work, we performed systematic calculations of the reaction intermediate for the ORR on the Fe-$\mathrm{N_{4}}$-C and Co-$\mathrm{N_{4}}$-C systems and calculated the overpotential for the ORR by using different XC functionals, and investigate how the choice of the XC functional impacts the predicted ORR activities of these systems regardless of the solvation effect.

\section{Methods}
\label{sec:methods}
\subsection{Computational Details}
\label{sec:computational_details}
All the calculations were performed using the projector augmented wave (PAW)\cite{blochl1994projector} method as implemented in the \textsc{Quantum-ESPRESSO} code \cite{giannozzi2009quantum}.
The PAW potentials were adopted from the \textsc{PSLIBRARY} \cite{dal2014pseudopotentials} version 1.0.0.
Wave functions and augmentation charge density were expanded in terms of a plane-wave basis set with the kinetic energy cutoffs of 80 and 800 Ry, respectively.
The Marzari-Vanderbilt cold smearing \cite{marzari1999thermal} with a smearing width of 0.02 Ry was used to treat the Fermi level.
In this work, we compare the performance of different XC functionals within GGA, specifically PBE \cite{perdew1996generalized} and revised PBE (RPBE) \cite{hammer1999improved} functionals.
Grimme's semiempirical dispersion correction (DFT-D3) \cite{grimme2010consistent} is also include for both functionals namely, PBE+D3 and RPBE+D3 to further investigate the importance of the dispersion force in the systems considered in this work.
Finally, we include the Bayesian error estimations functional with nonlocal correlation (BEEF-vdW) \cite{wellendorff2012density} to further assess the accuracy of the GGA functionals.
The BEEF-vdW is designed specifically to address vdW forces reasonably well while maintaining an accurate description of chemisorption energies of molecule on surface.
Moreover, the BEEF-vdW itself has an error estimate built in, to not only yield an accurate energy difference in condensed matter studies, but also estimate the errors on computed quantities.
Therefore, BEEF-vdW serves as a benchmark and a baseline for comparison in this study.
To employ the error estimation capabilities in the BEEF-vdW, a probability distribution for model parameters for the exchange and correlation is randomly sampled through an ensemble of density functionals generated around the BEEF-vdW \cite{wellendorff2012density,medford2014assessing}.
In this work, we generated an ensemble of 2000 functionals for each adsorption system, isolated surface and adsorbate considered, to obtain a distribution of the adsorption energies ($E_{\mathrm{ads}}$).
Then, the standard deviation of the ensemble for $E_{\mathrm{ads}}$ around the BEEF-vdW value was used to estimate the error for each adsorption system.
In addition to $E_{\mathrm{ads}}$, the standard deviation was also used to estimate the uncertainty in the limiting potential and overpotential.
We note that we did \textit{not} scale the error in such a way that it reproduces that for the benchmark adsorption energies, unlike Ref.~\cite{medford2014assessing}.
Thus, our estimated error for both $E_{\mathrm{ads}}$ and limiting potential/overpotential can be overestimated.
%

\begin{figure}[h]
   \includegraphics[width=0.9\columnwidth]{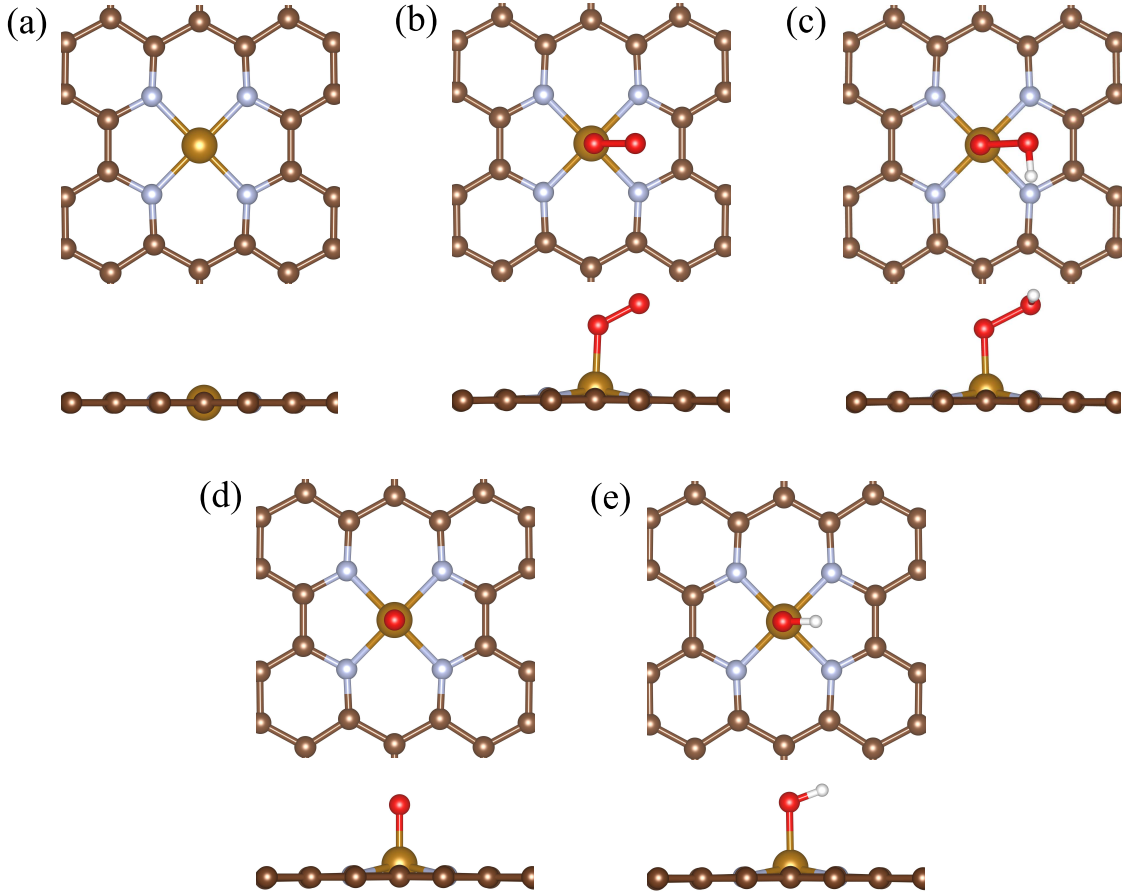}
    \caption{\label{fig:optimised} (a) The structure of a clean TM-N$_4$-C active site (TM is Fe or Co) Adsorption structures of (b) $\mathrm{^{\ast}O_{2}}$, (c) $\mathrm{^{\ast}OOH}$, (d) $\mathrm{^{\ast}O}$ and (e) $\mathrm{^{\ast}OH}$ on TM-N$_\mathrm{4}$. The gold, brown, red, silver and white atoms represent transition metal (Fe or Co), C, O, N and H, respectively.}
\end{figure}

%
Throughout this study, we used PAW potentials generated using the PBE functional.
Detailed results of the optimized lattice constant and the adsorption energies of the ORR intermediates obtained using RPBE and RPBE+D3 functionals with PAW potentials generated using the RPBE functionals are included in Supplemental Materials \cite{SM} for further reference.
We note that it was suggested \cite{kirchhoff2021assessment} that hybrid functions such as PBE0 \cite{perdew1996rationale,adamo1999toward} and HSE06 \cite{krukau2006influence} give more accurate results for ORR on nitrogen-doped graphene than GGA or SCAN meta-GGA\cite{sun2015strongly}, as compared with those obtained using the highly accurate coupled cluster theory.
Liu \textit{et al}. \cite{liu2022insights} used various XC functionals including HSE06 and SCAN and reported that the absolute value of the magnetic moment, which is suggested to be a good descriptor for ORR on Fe-N-C catalysts, varies depending on the functional, but the trend is the same (the authors do not report the reaction free energies with different functionals).
In this study, we limit ourselves to the GGA level of theory, in order to systematically and comprehensively study the roles of the XC functional and dispersion correction on not only the energetics, but also the vibrational contributions to the reaction free energies.
\begin{table}[h]
\caption{\label{tab:graphene_lattice_constant} Optimized lattice constant for pristine graphene obtained using each functionals compared to experimental value \cite{baskin1955lattice}. For RPBE and RPBE+D3 functionals, the optimized lattice constant are obtained using the PBE PAW potential.}
\begin{ruledtabular}
\begin{tabular}{lc}
Functional & Lattice constant (\AA) \\ \hline
Experimental & 2.460 \\
PBE & 2.467 \\
PBE+D3 & 2.467 \\
RPBE & 2.477 \\
RPBE+D3 & 2.479 \\
BEEF-vdW & 2.465 \\
\end{tabular}
\end{ruledtabular}
\end{table}
We considered Fe-N$_4$ and Co-N$_4$ moieties embedded in graphene.
The lattice constant for graphene was optimized in the primitive hexagonal unit cell using a 16$\times$16 $\mathbf{k}$-point grid for each functional.
The DFT-optimized lattice constants are given in the Table ~\ref{tab:graphene_lattice_constant}.
For the calculation of the ORR intermediates, the Fe-$\mathrm{N_{4}}$-C and Co-$\mathrm{N_{4}}$-C models was constructed using a (5$\times$5) surface unit cell of graphene, in which we removed two carbon atoms to anchor a single Fe/Co atom and replaced four-coordinated carbon atoms surrounding it with nitrogen atom [Fig.~\ref{fig:optimised}(a)].
To eliminate the spurious electrostatic interaction with the periodic images, the effective screening medium method \cite{otani2006first,hamada2009green} was employed, along with the vacuum spacing of 20~\AA.
The geometries of all the reaction intermediate ($\mathrm{^{\ast}O_{2}}$, $\mathrm{^{\ast}OOH}$, $\mathrm{^{\ast}O}$, and $\mathrm{^{\ast}OH}$) on both Fe-$\mathrm{N_{4}}$-C and Co-$\mathrm{N_{4}}$-C moeties were fully optimized until the residual forces on the constituent atoms became smaller than $10^{-4}$ Ry/Bohr ($2.57 \times 10^{-3}$ eV/\AA).
A 4$\times$4 $\mathbf{k}$-point grid was found to give reasonably accurate results (see Supplemental Materials \cite{SM}) and 8$\times$8 $\mathbf{k}$-point grid was used for the electronic structure analyses.
The adsorption energy of the intermediate is defined by
\begin{equation}
E_{\mathrm{ads}} = E_{\mathrm{S/A}} - E_{\mathrm{S}} - E_{\mathrm{A}},
\label{eqn:eint}
\end{equation}
where
$E_{\mathrm{S/A}}$, $E_{\mathrm{S}}$, and $E_{\mathrm{A}}$, are the total energies of combined system, isolated surface structure, and isolated adsorbate, respectively.
All energies above were obtained under the same parameter settings.
\subsection{Free energy calculation}
\label{sec:free energy}

The change in adsorption energy with an applied electrode potential was calculated based on the CHE model proposed by N{\o}rskov and co-workers \cite{norskov2004origin}.
In this study, we focus on the associative mechanism of the ORR process as it has been reported that it is energetically more favorable and that the activation barrier for dissociative O$_2$ adsorption is considerably high (1.16 eV)\cite{wang2019self} in the cases of the Fe-N$_4$-C and Co-N$_4$-C active sites.
%
%
The elementary steps for ORR along the four-electron associative mechanisms can be written as follows:
\begin{align}
   &\begin{aligned}
         {^{\ast}} + \mathrm{O_{2}} + \mathrm{H}^{+} + e^{-} \rightarrow                     {^{\ast}}\mathrm{OOH}\\
   \end{aligned}\\
   &\begin{aligned}
        {^{\ast}}{\mathrm{OOH}} + \mathrm{H}^{+} + e^{-} \rightarrow {^{\ast}{\mathrm{O}}} + \mathrm{H_{2}{O}} \\
   \end{aligned}\\
   &\begin{aligned}
        {^{\ast}}{\mathrm{O}} + \mathrm{H}^{+} + e^{-} \rightarrow {^{\ast}{\mathrm{OH}}} \\
   \end{aligned}\\
    &\begin{aligned}
        {^{\ast}}{\mathrm{OH}} + \mathrm{H}^{+} + e^{-} \rightarrow {^{\ast}} + \mathrm{H_{2}{O}} \\
   \end{aligned}
\end{align}
where $^{\ast}$ denotes the adsorption site.
The DFT reaction energies of the $^{\ast}\mathrm{OOH}$, $^{\ast}\mathrm{O}$ and $^{\ast}\mathrm{OH}$ are then defined relative to $\mathrm{H_{2}{O}}$ (liquid phase) and $\mathrm{H_{2}}$ (gas phase), to avoid the calculation of the $\mathrm{O_{2}}$ molecule or the radical of $\mathrm{OOH}$ and $\mathrm{OH}$, which is notoriously difficult to describe within the semilocal approximation to DFT \cite{norskov2004origin} as:
\begin{align}
   &\begin{aligned}
         {\Delta{E}_{\mathrm{OOH}}} = {E_{\ast{\mathrm{OOH}}}} - {E_{\ast}} - ({2E_{\mathrm{H_{2}O(l)}}} - {\frac{3}{2}}E_{\mathrm{H_{2}(g)}})
   \end{aligned}\\
   &\begin{aligned}
        {\Delta{E}_{\mathrm{O}}} = {E_{\ast{\mathrm{O}}}} - {E_{\ast}} - ({E_{\mathrm{H_{2}O(l)}}} - {E_{\mathrm{H_{2}(g)}}})\\
   \end{aligned}\\
    &\begin{aligned}
        {\Delta{E}_{\mathrm{OH}}} = {E_{\ast{\mathrm{OH}}}} - {E_{\ast}} - ({E_{\mathrm{H_{2}O(l)}}} - {\frac{1}{2}}E_{\mathrm{H_{2}(g)}})\\
   \end{aligned}
\end{align}
where $E_{\ast}$ is the total energy of the surface without adsorbates, $E_{\ast{\mathrm{OOH}}}$, $E_{\ast{\mathrm{O}}}$ and $E_{\ast{\mathrm{OH}}}$ are total energies of the surface with adsorbates ($^{*}\mathrm{OOH}$, $^{*}\mathrm{O}$ and $^{*}\mathrm{OH}$, respectively) bound to the surface, $E_{\mathrm{H_{2}O(g)}}$ and $E_{\mathrm{H_{2}(g)}}$ are the total energies of $\mathrm{H_{2}O}$ and $\mathrm{H_{2}}$ molecules in the gas phase, respectively.
The energy of H$_2$O in the liquid phase ($E_{\mathrm{H2O(l)}}$) is calculated at 0.035 bar (i.e. $E_{\mathrm{H_2O(l)}}$ = $E_{\mathrm{H_2O(g)}}$ + $k_{\mathrm{B}}T\ln0.035$), which correspond to equilibrium pressure of $\mathrm{H_{2}O}$ at 298.15 K, and therefore this state corresponds to that of liquid water \cite{weast1971handbook}.
The DFT formation energies for the ORR intermediates are converted to the adsorption free energy by including corrections for the change in zero-point energy $\Delta \mathrm{ZPE}$ and entropy $T \Delta S$ at 298.15 K estimated using finite displacement method to calculate the vibrational frequencies of the adsorbates through the use of the atomic simulation environment (ASE) \cite{larsen2017atomic} and $\Delta G_{\mathrm{pH}}$ are the pH values (pH was defined as 0 for acidic medium) as
\begin{equation}
 \Delta{G} = \Delta E_{\mathrm{DFT}} + \Delta\mathrm{ZPE} - T\Delta{S} + \Delta G_{\mathrm{pH}}.
\label{eqn:efg}
\end{equation}
This allows us to define the adsorption free energies of $^{*}\mathrm{OOH}$, $^{*}\mathrm{O}$ and $^{*}\mathrm{OH}$ (i.e. along associative mechanism) relative to $\mathrm{H_{2}O}$ and $\mathrm{H_{2}}$
\begin{align}
   &\begin{aligned}
         {^{\ast}} + {\mathrm{H_{2}O}} {\rightarrow} {^{*}{\mathrm{O}}} + {\mathrm{H_{2}}}, & &{\Delta{G}(^{*}{\mathrm{O})}}
   \end{aligned}\\
   &\begin{aligned}
        {^{\ast}} + {\mathrm{H_{2}O}} {\rightarrow} {^{*}{\mathrm{OH}}} + {\frac{1}{2}\mathrm{H_{2}}}, & &{\Delta{G}(^{*}{\mathrm{OH})}}
   \end{aligned}\\
    &\begin{aligned}
        {^{\ast}} + 2{\mathrm{H_{2}O}} {\rightarrow} {^{\ast}{\mathrm{OOH}}} + {\frac{3}{2}\mathrm{H_{2}}}, & &{\Delta{G}(^{\ast}{\mathrm{OOH})}}
   \end{aligned}
\end{align}
In the case of the adsorption free energy of O$_{2}$ ($^{\ast}$ + $\mathrm{O_{2}}$ $\rightarrow$ $^{\ast}{\mathrm{O_{2}}}$), the adsorption energy is defined relative to an $\mathrm{O}_2$ molecule in the gas phase ($\mathrm{O_{2}}$(g)).
By setting the reversible hydrogen electrode (RHE) as the reference electrode, the effect of the applied potential ($U$ $\mathrm{V_{RHE}}$) on the Gibbs free energy can be approximated by
\begin{equation}
    \Delta G_{U} = \Delta G - neU,
\label{eqn:elec}
\end{equation}
where $n$ is the number of electrons transferred in each consecutive step and $e$ is the elementary charge of an electron.
The Gibbs free energy of the four elementary reactions steps in the associative mechanism (at $U$ = 0 $\mathrm{V_{RHE}}$) can be written as
\begin{align}
   &\begin{aligned}
         \Delta{G_{1}} &= \Delta{G(^{\ast}{\mathrm{OOH}})} - \Delta{G_{\mathrm{ORR}}}
   \end{aligned}\\
   &\begin{aligned}
        \Delta{G_{2}} &= \Delta{G(^{\ast}{\mathrm{O}})} - \Delta{G(^{\ast}{\mathrm{OOH}})}
   \end{aligned}\\
    &\begin{aligned}
        \Delta{G_{3}} &= \Delta{G(^{\ast}{\mathrm{OH}})} - \Delta{G(^{\ast}{\mathrm{O}})}
   \end{aligned}\\
   &\begin{aligned}
        \Delta{G_{4}} &= - \Delta{G(^{\ast}{\mathrm{OH}})}
   \end{aligned}
\end{align}
where
$\Delta{G_{\mathrm{ORR}}}$ is the overall formation free energy value of ORR through the relation (2$\mathrm{H_{2}}$ + $\mathrm{O_{2}}$ $\rightarrow$ 2$\mathrm{H_{2}{O}}$), defined experimentally ($\Delta{G_{\mathrm{ORR}}}$ = $-$4.92 eV) at 298.15 $\mathrm{K}$ with pressure of $\mathrm{O_{2}}$ and $\mathrm{H_{2}}$ of 1 bar \cite{rossmeisl2005electrolysis}.
With this value, the thermodynamic equilibrium potential will be 1.23 $\mathrm{V_{RHE}}$.
Based on the calculations described above, the free energy diagram along the reaction path considered can be constructed.
The limiting-potential (${U_{\mathrm{L}}}$) is estimated from Eq. (\ref{eqn:lim}) by determining the minimum free energy change ($\Delta{G_{\mathrm{min}}}$) among all the electrochemical steps.
\begin{equation}
    U_{\mathrm{L}} = \mathrm{min}{\{{\Delta{G_{1}},{\Delta{G_{2}}},{\Delta{G_{3}},{\Delta{G_{4}}\}}/{e}}}}
\label{eqn:lim}
\end{equation}
In the CHE model, the ${U_{\mathrm{L}}}$ is also known as potential at which all the reaction steps are exothermic and by this definition, it can be used to compare with the experimental half-wave potential ($\mathrm{E_{\frac{1}{2}}}$) value \cite{dickens2019insights}.
Accordingly, the overpotential ($\eta_{ORR}$) can be further derived from the (${U_{\mathrm{L}}}$) as Eq. (\ref{eqn:over}), in which a lower overpotential implies a better ORR activity.
\begin{equation}
    \eta_{\mathrm{ORR}} = 1.23 - U{\mathrm{_L}}
    \label{eqn:over}
\end{equation}
\section{Result and Discussion}
\label{sec:results}
\subsection{Adsorption energies for the ORR intermediates}
\label{sec:binding}
\begin{figure}[h]
   \includegraphics[width=1.0\columnwidth]{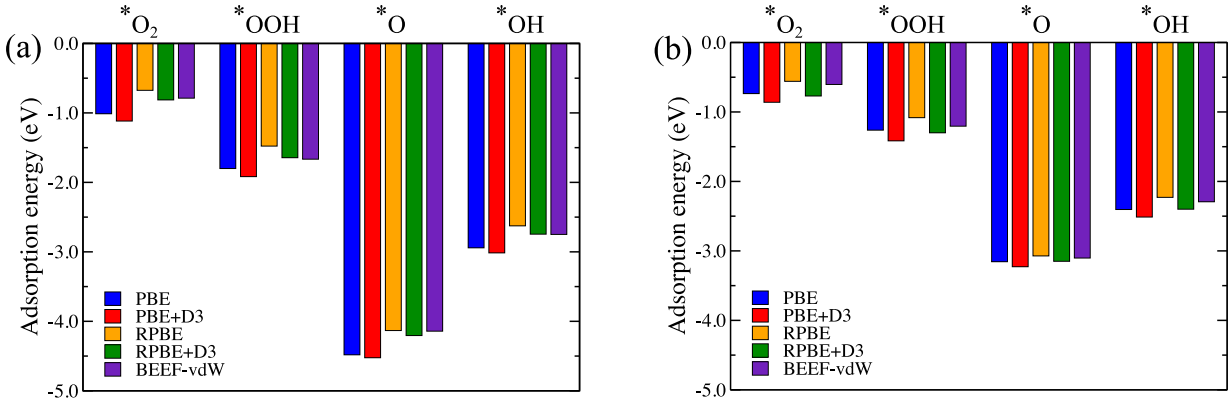}
    \caption{\label{fig:ads_energy} Calculated adsorption energies for the ORR intermediates on (a) Fe-$\mathrm{N_{4}}$-C and (b) Co-$\mathrm{N_{4}}$-C with difference functional considered.}
\end{figure}

We first optimized the ORR intermediates ($\mathrm{^{\ast}O_{2}}$, $\mathrm{^{\ast}OOH}$, $\mathrm{^{\ast}O}$, and $\mathrm{^{\ast}OH}$) on  Fe-$\mathrm{N_{4}}$-C and Co-$\mathrm{N_{4}}$-C active sites, and calculated their adsorption energies with PBE, PBE+D3, RPBE, RPBE+D3 and BEEF-vdW functionals (Fig.~\ref{fig:ads_energy}).
The calculated energies are summarizes in Table ~\ref{tab:int_eng}.
\begin{table}[h]
\caption{\label{tab:int_eng} Adsorption energies for the ORR adsorbates on the Fe-$\mathrm{N_{4}}$-C and Co-$\mathrm{N_{4}}$-C active sites with different functionals considered. The unit of energy is eV.}
\begin{ruledtabular}
\begin{tabular}{cccccc}
Intermediate species & PBE & PBE+D3 & RPBE & RPBE+D3 & BEEF-vdW\\
\hline
\multicolumn{6}{c}{Fe-$\mathrm{N_{4}}$-C}\\
$\mathrm{{^\ast}O_{2}}$ & $-$1.01 & $-$1.11 & $-$0.67 & $-$0.81 & $-$0.79 $\pm$ 0.3\\
$\mathrm{{^\ast}OOH}$ & $-$1.80 & $-$1.91 & $-$1.48 & $-$1.64 & $-$1.67 $\pm$ 0.3\\
$\mathrm{{^\ast}O}$ & $-$4.48 & $-$4.52  & $-$4.13 & $-$4.20 & $-$4.14 $\pm$ 0.2\\
$\mathrm{{^\ast}OH}$ & $-$2.94 & $-$3.01 & $-$2.62 & $-$2.75 & $-$2.75 $\pm$ 0.2\\
\multicolumn{6}{c}{Co-$\mathrm{N_{4}}$-C}\\
$\mathrm{{^\ast}O_{2}}$ & $-$0.73 & $-$0.86 & $-$0.56 & $-$0.77 & $-$0.60 $\pm$ 0.2\\
$\mathrm{{^\ast}OOH}$ & $-$1.26 & $-$1.41 & $-$1.08 & $-$1.33 & $-$1.20 $\pm$ 0.2\\
$\mathrm{{^\ast}O}$ & $-$3.15 & $-$3.23 & $-$3.07 & $-$3.20 & $-$3.10 $\pm$ 0.1\\
$\mathrm{{^\ast}OH}$ & $-$2.40 & $-$2.51 & $-$2.23 & $-$2.43 & $-$2.29 $\pm$ 0.1\\
\end{tabular}
\end{ruledtabular}
\end{table}
The present PBE results on both active sites are in a good agreement with the previous works by Kattel \textit{et al}. \cite{kattel2014density,kattel2013catalytic}.
In general, the adsorption energies obtained using PBE are larger in magnitude, and by the inclusion of dispersion correction, PBE+D3 gives adsorption energies larger in magnitude than those obtained using other functionals, and the energy differences are more significant in the case of Fe-$\mathrm{N_{4}}$-C active site.
In contrast, RPBE predicts more repulsive interactions and resulted in more positive interaction energies on both active sites.
However, RPBE+D3 improves the adsorption energies, which are larger in magnitude than those by RPBE, and the values are in good agreement with those obtained by using BEEF-vdW.
The results indicate that the contribution of dispersion interaction is significant within these two systems, and the inclusion of the dispersion correction in RPBE+D3 leads to the adsorption energies of all the ORR intermediates with similar accuracy to the BEEF-vdW.
We note that the estimated error in the BEEF-vdW adsorption energies is relatively large (0.2 - 0.3 eV for Fe-$\mathrm{N_{4}}$-C and 0.1 - 0.2 eV for Co-$\mathrm{N_{4}}$-C), but agrees well with the previously reported one for N-doped graphene \cite{kirchhoff2021assessment}.
We also note that PBE tends to overestimate the adsorption energy of chemisorption species, partly because the exchange in the PBE functional tends to lead too attractive interaction in molecular systems \cite{hammer1999improved}.
In addition, PBE can show spurious attractive interaction for weakly interacting systems from the exchange only.
The detail discussion can be found in Ref.~\onlinecite{abidin2022interaction,hamada2010interaction,hamada2012adsorption}.
\subsection{Reaction mechanism of the ORR}
\label{sec:free}
\begin{figure}[h]
   \includegraphics[width=1.0\columnwidth]{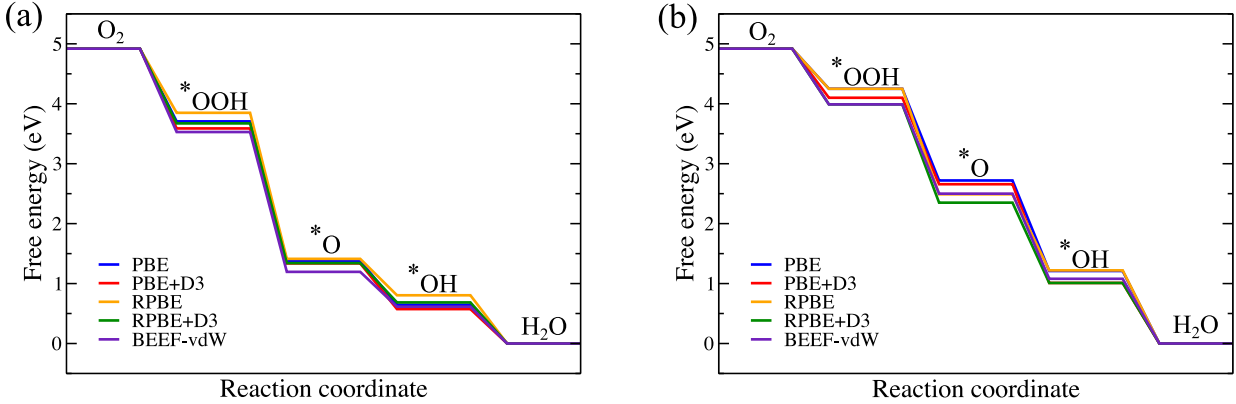}
    \caption{\label{fig:free} Calculated free energy diagrams of ORR along the associative mechanism on (a) Fe-$\mathrm{N_{4}}$-C and (b) Co-$\mathrm{N_{4}}$-C catalysts at ($U$ = 0 V$_{\mathrm{RHE}}$).}
\end{figure}

We next evaluated the catalytic activities of the Fe-$\mathrm{N_{4}}$-C and Co-$\mathrm{N_{4}}$-C for the ORR with different functionals.
The free energy diagrams for ORR along the four-electron associative mechanism on the Fe-$\mathrm{N_{4}}$-C and the Co-$\mathrm{N_{4}}$-C catalysts are calculated and displayed in Fig. \ref{fig:free}.
The calculated adsorption free energies and details theoretical limiting potentials and overpotentials on the Fe-$\mathrm{N_{4}}$-C and the Co-$\mathrm{N_{4}}$-C catalysts are summarized in Tables ~\ref{tab:ads-free} and \ref{tab:limit}, respectively.
\begin{table}[h]
\caption{\label{tab:ads-free} Calculated adsorption free energies for $\mathrm{{^\ast}OOH}$, $\mathrm{{^\ast}O}$ and $\mathrm{{^\ast}OH}$ on Fe-$\mathrm{N_{4}}$-C and Co-$\mathrm{N_{4}}$-C active sites with different functionals considered. The unit of energy is eV.}
\begin{ruledtabular}
\begin{tabular}{cccccc}
 & PBE & PBE+D3 & RPBE & RPBE+D3 & BEEF-vdW\\
\hline
\multicolumn{6}{c}{Fe-$\mathrm{N_{4}}$-C}\\
${\Delta{G}_{{^\ast}\mathrm{OOH}}}$ & 3.70 & 3.59 & 3.85 & 3.66 & 3.52\\
${\Delta{G}_{{^\ast}\mathrm{O}}}$ & 1.41 & 1.37 & 1.42 & 1.37 & 1.23\\
${\Delta{G}_{{^\ast}\mathrm{OH}}}$ & 0.67 & 0.60 & 0.81 & 0.71 & 0.64\\
\multicolumn{6}{c}{Co-$\mathrm{N_{4}}$-C}\\
${\Delta{G}_{{^\ast}\mathrm{OOH}}}$ & 4.25 & 4.09 & 4.25 & 3.99 & 3.99\\
${\Delta{G}_{{^\ast}\mathrm{O}}}$ & 2.72 & 2.62 & 2.49 & 2.35 & 2.50\\
${\Delta{G}_{{^\ast}\mathrm{OH}}}$ & 1.21 & 1.10 & 1.22 & 1.01 & 1.09\\
\end{tabular}
\end{ruledtabular}
\end{table}
\begin{table}[h]
\caption{\label{tab:limit} Calculated theoretical limiting potential (${U_{\mathrm{L}}}$) and overpotential ($\eta_{\mathrm{ORR}}$) for the Fe-$\mathrm{N_{4}}$-C and Co-$\mathrm{N_{4}}$-C catalysts.
The potential determining steps (PDS) for the Fe-$\mathrm{N_{4}}$-C are H$_2$O formation with PBE, PBE+D3, and OH formation with  RPBE, RPBE+D3 and BEEF-vdW. For the Co-$\mathrm{N_{4}}$-C, all functionals predict the OOH formation as PDS.
The unit of potential is volt.}
\begin{ruledtabular}
\begin{tabular}{cccccc}
 & PBE & PBE+D3 & RPBE & RPBE+D3 & BEEF-vdW  \\
\hline
\multicolumn{6}{c}{Fe-$\mathrm{N_{4}}$-C}\\
Limiting potential  & 0.67 & 0.60 & 0.61 & 0.66 & 0.59 $\pm$ 0.4\\
Overpotential & 0.56 & 0.63 & 0.62 & 0.57 & 0.64 $\pm$ 0.4\\
\multicolumn{6}{c}{Co-$\mathrm{N_{4}}$-C}\\
Limiting potential  & 0.67 & 0.82 & 0.67 & 0.93 & 0.93 $\pm$ 0.3\\
Overpotential & 0.56 & 0.41 & 0.56 & 0.30 & 0.30 $\pm$ 0.3\\
\end{tabular}
\end{ruledtabular}
\end{table}
\begin{figure}[h]
   \includegraphics[width=1.0\columnwidth]{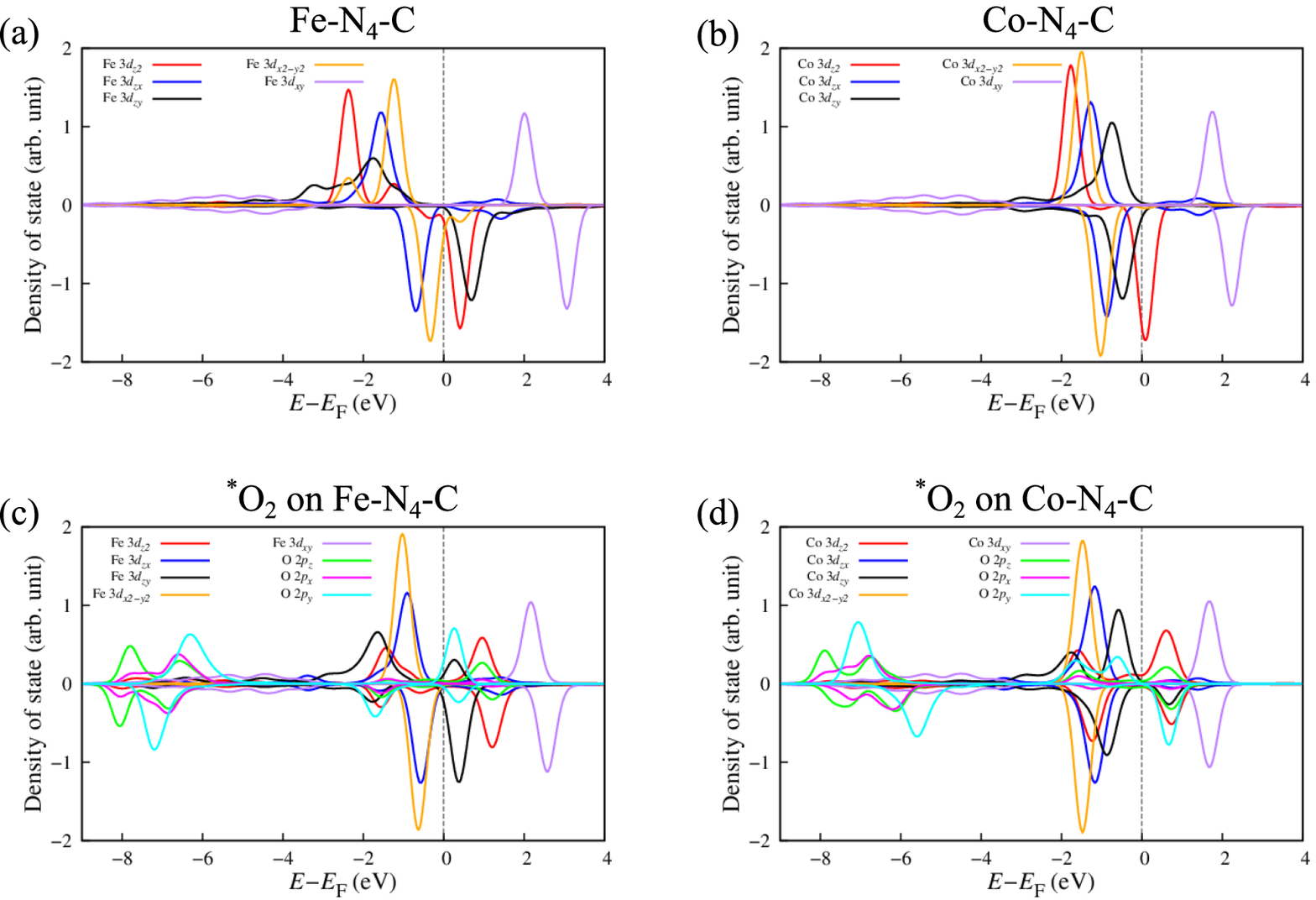}
    \caption{\label{fig:pdos} The partial density of state for Fe 3$d$, Co 3$d$ and O 2$p$ of Fe-$\mathrm{N_{4}}$-C and Co-$\mathrm{N_{4}}$-C active sites (a), (b) without and (c), (d) with $\mathrm{^{\ast}O_{2}}$ adsorbate. The origin of the energy is set to the Fermi level ($E_{\mathrm{F}}$). The positive densities of states are for the spin up channel, while negative, the spin down one.}
\end{figure}
Our calculated ORR free energy diagrams for the associative mechanism (Fig. \ref{fig:free}) show that RPBE and RPBE+D3 functionals predict a similar potential determining step (PDS) with BEEF-vdW, which is the O protonation to form OH, while both PBE and PBE+D3 predict the formation of H$_2$O (last reaction steps) on the Fe-$\mathrm{N_{4}}$-C.
On the other hand, all the functionals predict the OOH formation (initial reaction steps) as the PDS on the Co-$\mathrm{N_{4}}$-C catalyst.
The difference in the estimated PDSs in the Fe-$\mathrm{N_{4}}$-C catalyst originates from the significant difference in the ORR intermediate adsorption energies predicted by the PBE and PBE+D3 functionals.
The ${U_{\mathrm{L}}}$ values for Fe-$\mathrm{N_{4}}$-C predicted by PBE+D3 and RPBE are in good agreement with that by BEEF-vdW with a difference of $\sim$ 0.02 V, while PBE and RPBE+D3 is slightly overestimate it by $\sim$ 0.08 V.
It should be stressed that all the functionals predicts ${U_{\mathrm{L}}}$'s  for Fe-$\mathrm{N_{4}}$-C, which seemingly agree well with that from BEEF-vdW.
However, PBE and PBE+D3 values are obtained for a different PDS, i.e., different mechanisms.
On the other hand, in the case of Co-$\mathrm{N_{4}}$-C, only RPBE+D3 predicts ${U_{\mathrm{L}}}$, which is comparable to that from BEEF-vdW, while PBE, PBE+D3, and RPBE underestimate it by $\sim$ 0.25, $\sim$ 0.10, and $\sim$ 0.26 V, respectively.
We also estimated the $\eta_{\mathrm{ORR}}$ by using Eq.~\ref{eqn:over} (Table \ref{tab:limit}).
The error bars for BEEF-vdW indicate a relatively large uncertainty (0.4 V) for the $\eta_{\mathrm{ORR}}$ on both active sites.
Nevertheless, only RPBE+D3 consistently estimated a similar PDS and comparable $\eta_{\mathrm{ORR}}$ to BEEF-vdW.
The inconsistent results on both catalysts  obtained using PBE and PBE+D3 may be due to the too attractive exchange in PBE, as discussed above.
With respect to the $\eta_{\mathrm{ORR}}$ values (Table \ref{tab:limit}), PBE+D3, RPBE+D3, and BEEF-vdW predict that the Co-$\mathrm{N_{4}}$-C catalyst is more reactive (lower $\eta_{\mathrm{ORR}}$ value), while PBE and RPBE estimate the catalytic performance of the Fe-$\mathrm{N_{4}}$-C and the Co-$\mathrm{N_{4}}$-C are comparable.
It is noted that our results in Table \ref{tab:limit} are limited to the vacuum condition.
It is also noted that we estimated the adsorption free energies using RPBE+D3 with the on-site Coulomb interaction ($U$) and report in Supplemental Materials \cite{SM}, which are consistent with those reported in Ref.~\cite{wang2019self}.
However, calculated adsorption energies and potential determining steps can depend on the choice of the $U$ value, which calls for further investigation.
\subsection{Electronic structures}
To gain insight into their catalytic activities, we calculated the densities of states for clean and $\mathrm{{^\ast}O_{2}}$ adsorbed Fe-$\mathrm{N_{4}}$-C and Co-$\mathrm{N_{4}}$-C sites using RPBE+D3 functional.
We found that a distinct spin polarization at the Fe-$\mathrm{N_{4}}$-C site compared to the Co-$\mathrm{N_{4}}$-C sites, which can be also confirmed by a larger local magnetic moment (1.61 $\mu_{\mathrm{B}}$ and 0.68 $\mu_{\mathrm{B}}$ for the former and the latter, respectively).
These results further explain stronger binding strength for all ORR intermediates that we obtained for the Fe-$\mathrm{N_{4}}$-C as compared to Co-$\mathrm{N_{4}}$-C site (Table \ref{tab:int_eng}).
Besides, the spin down channels of $3d_{z^2}$ and $3d_{zy}$ in the Fe-$\mathrm{N_{4}}$-C are unoccupied and located above the Fermi level, while the Co-$\mathrm{N_{4}}$-C has the half-filled $3d_{z^2}$ spin down channel cross the Fermi level.
This indicates that the Co-N$_{4}$-C active site has a half-metallic nature, and has less opportunities to form a bond with O.
Therefore, the Fe-$\mathrm{N_{4}}$-C site can form a stronger Fe$-$O $\sigma$ bond compare to Co$-$O \cite{tsuda2004cyclohexane,tsuda2004spin,tsuda2005comparative}.
This behavior can be related to the Sabatier principle in which the strong bonding between active site and reactant effectively inhibits further reaction and results in decreasing catalytic activity \cite{sabatier1920catalyse,che2013nobel}.
As expected, the moderate spin polarization and half-metallic properties shown by the Co-$\mathrm{N_{4}}$-C site ensures the moderate binding strength of the main ORR intermediates, and is responsible for the higher ORR catalytic activity \cite{sun2019itinerant}.
\subsection{Effect of OH termination of the active Fe/Co site}
\label{sec:OH-terminated}
\begin{figure}[h]
   \includegraphics[width=1.0\columnwidth]{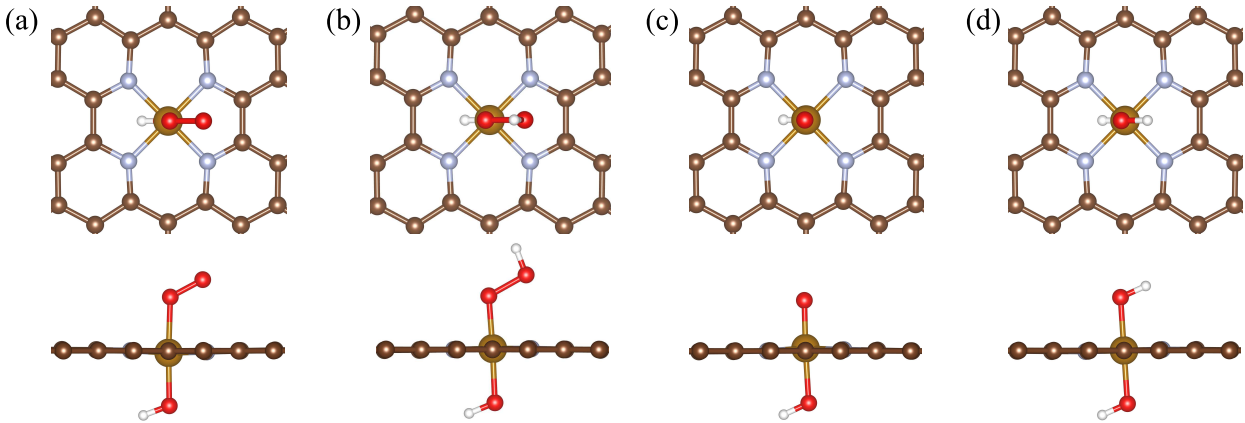}
    \caption{\label{fig:oh-optimised} The adsorption structures of (a) $\mathrm{^{\ast}O_{2}}$, (b) $\mathrm{^{\ast}OOH}$, (c) $\mathrm{^{\ast}O}$ and (d) $\mathrm{^{\ast}OH}$ on OH-terminated TM(OH)-N$_\mathrm{4}$ active site (TM is Fe or Co). The gold, brown, red, silver and white atoms represent transition metal (Fe or Co), C, O, N and H, respectively.}
\end{figure}

In a recent theoretical work by Wang \textit{et al}. \cite{wang2019self}, it was proposed that the intrinsic intermediate $\mathrm{{^\ast}OH}$ is co-adsorbed on the single-metal-atom active site, and is the origin of the improved activity of the ORR through the associative mechanism on the TM-$\mathrm{N_{4}}$-C catalyst.
The presence of the OH species was also confirmed by the experiment \cite{jia2015experimental}.
Following the aforementioned studies, we further investigate the effect of OH-termination on the Fe-$\mathrm{N_{4}}$-C and Co-$\mathrm{N_{4}}$-C active sites without the solvent effect (Fig. \ref{fig:oh-optimised}).
Here we employed RPBE+D3 functional, which predicts consistent results with BEEF-vdW and compares with the results obtained using PBE+D3.
We also include PBE in this comparison, because it was used in Ref.~\onlinecite{wang2019self}.
The calculated energies are summarizes in Table \ref{tab:terminted-ads}
\begin{table}[h]
\caption{\label{tab:terminted-ads} Calculated adsorption free energies for $\mathrm{{^\ast}OOH}$, $\mathrm{{^\ast}O}$ and $\mathrm{{^\ast}OH}$ on Fe-$\mathrm{N_{4}}$-C and Co-$\mathrm{N_{4}}$-C active sites with and without OH termination of the Fe/Co sites. The PBE, PBE+D3, and RPBE+D3 functionals was used and the unit of energy is eV.}
\begin{ruledtabular}
\begin{tabular}{ccccc}
 & Fe-$\mathrm{N_{4}}$-C  & Fe(OH)-$\mathrm{N_{4}}$-C  & Co-$\mathrm{N_{4}}$-C & Co(OH)-$\mathrm{N_{4}}$-C\\
\hline
\multicolumn{5}{c}{PBE}\\
${\Delta{G}_{{^\ast}\mathrm{OOH}}}$  & 3.70 & 4.11 & 4.25 & 4.22\\
${\Delta{G}_{{^\ast}\mathrm{O}}}$    & 1.41 & 2.17 & 2.72 & 2.80\\
${\Delta{G}_{{^\ast}\mathrm{OH}}}$   & 0.67 & 0.99 & 1.21 & 1.11\\
\multicolumn{5}{c}{PBE+D3}\\
${\Delta{G}_{{^\ast}\mathrm{OOH}}}$  & 3.59 & 3.94 & 4.10 & 4.05\\
${\Delta{G}_{{^\ast}\mathrm{O}}}$    & 1.37 & 2.07 & 2.65 & 2.70\\
${\Delta{G}_{{^\ast}\mathrm{OH}}}$   & 0.60 & 0.85 & 1.10 & 0.98\\
\multicolumn{5}{c}{RPBE+D3}\\
${\Delta{G}_{{^\ast}\mathrm{OOH}}}$  & 3.66 & 3.91 & 3.99 & 4.03\\
${\Delta{G}_{{^\ast}\mathrm{O}}}$  & 1.37 & 1.96 & 2.35 & 2.59\\
${\Delta{G}_{{^\ast}\mathrm{OH}}}$  & 0.71 & 0.84 & 1.01 & 0.96\\
\end{tabular}
\end{ruledtabular}
\end{table}
\begin{figure}[h]
   \includegraphics[width=1.0\columnwidth]{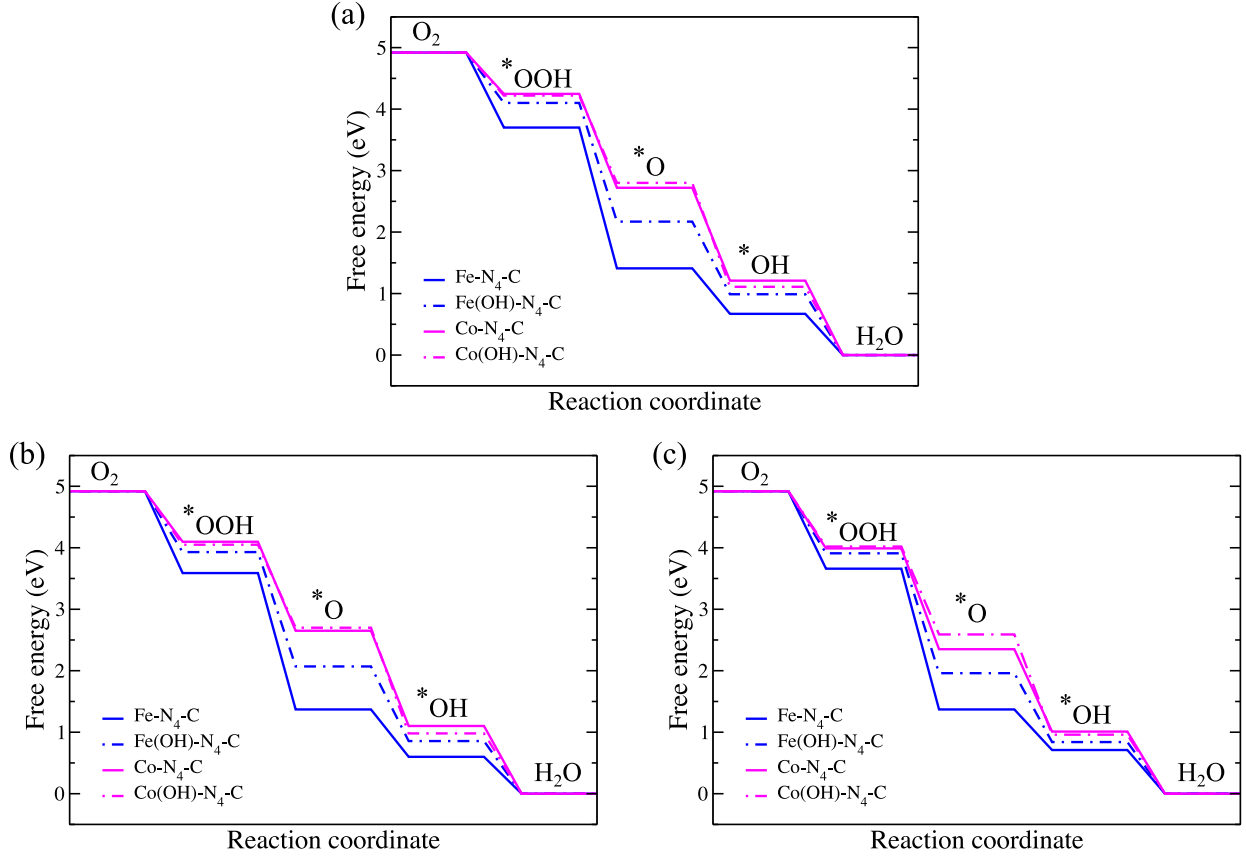}
    \caption{\label{fig:terminated-free} Free energy diagrams of ORR along associative pathway on the Fe-$\mathrm{N_{4}}$-C and Co-$\mathrm{N_{4}}$-C active sites with and without OH-termination obtained using (a) PBE, (b) PBE+D3, and (c) RPBE+D3 functionals.}
\end{figure}
Here, with the OH-termination of both Fe-N$_4$-C and Co-N$_4$-C active sites, PBE+ D3 and RPBE+D3 predict the formation of $\mathrm{H_{2}O}$ as the PDS, while PBE, the protonation of $\mathrm{O_{2}}$ to form $\mathrm{OOH}$ on the Fe-N$_4$-C.
In the case of Co-N$_4$-C, all the functionals predicts the $\mathrm{O_{2}}$ protonation to form $\mathrm{OOH}$ as the PDS.
We further estimated the $\eta_{\mathrm{ORR}}$ for the detailed comparison (Table \ref{tab:limit2}).
The $\eta_{\mathrm{ORR}}$ values for Fe-N$_4$-C predicted by PBE, PBE+D3 and RPBE+D3 show a noticeable improvement of the ORR catalytic activities (lowering of the overpotential) after the OH co-adsorbs on the Fe site by $\sim$ 0.14, $\sim$ 0.25, and $\sim$ 0.17 V, respectively.
On the other hand, on the Co site, no significant change can be found by using each functionals.
The slight improvement of the ORR activity in the Fe-N$_4$-C is mainly due to the formation of the $\sigma$-bonding between $2p_{z}$ and $3d_{z^2}$ orbitals upon termination of the $\mathrm{{^\ast}OH}$ on the Fe site (see Supplemental Materials \cite{SM}).
Accordingly, the proportion of $\sigma$-bonding state on the Fe(OH)-$\mathrm{N_{4}}$-C site decreased, and the interaction between the ORR intermediates is dominated by the $\pi$-bonding states (i.e., the filling degree of $3d_{zx}$ and $3d_{zy}$).
Therefore, the adsorption of ORR intermediates on the Fe(OH)-$\mathrm{N_{4}}$-C becomes weaker (see Supplemental Materials \cite{SM}), and the ORR performance is improved.
The same behavior is observed for the Co-$\mathrm{N_{4}}$-C with the OH-termination.
However, since all the orbitals of $3d_{z^2}$, $3d_{zx}$, and $3d_{zy}$  are almost filled after the $\mathrm{{^\ast}OH}$ co-adsorption at the Co site, and also due to the half-metallic nature of the Co-$\mathrm{N_{4}}$-C, no significant change is observed in the adsorption strength of the ORR intermediates as well as the ORR catalytic activity.
\begin{table}[h]
\caption{\label{tab:limit2} Calculated limiting potential (${U_{\mathrm{L}}}$) and overpotential ($\eta_{\mathrm{ORR}}$) for the Fe-$\mathrm{N_{4}}$-C and Co-$\mathrm{N_{4}}$-C with and without OH termination of the Fe/Co active sites. The PBE, PBE+D3, and RPBE+D3 functionals was used and the unit of potential is volt.
The potential determining steps (PDS) for the Fe-$\mathrm{N_{4}}$-C without OH termination are H$_2$O formation with PBE, PBE+D3, and OH formation with RPBE+D3, while with OH termination, the PDS are OOH formation with PBE, and H$_2$O formation with PBE+D3, RPBE+D3, respectively. For the Co-$\mathrm{N_{4}}$-C, all functionals predict the OOH formation as PDS, without and with OH termination.
}
\begin{ruledtabular}
\begin{tabular}{ccccc}
 & Fe-$\mathrm{N_{4}}$-C & Fe(OH)-$\mathrm{N_{4}}$-C & Co-$\mathrm{N_{4}}$-C & Co(OH)-$\mathrm{N_{4}}$-C\\
\hline
\multicolumn{5}{c}{PBE}\\
Limiting potential  & 0.67 & 0.81 & 0.67 & 0.69\\
Overpotential & 0.56 & 0.42 & 0.55 & 0.53\\
\multicolumn{5}{c}{PBE+D3}\\
Limiting potential  & 0.60 & 0.85 & 0.82 & 0.87\\
Overpotential & 0.63 & 0.37 & 0.41 & 0.36\\
\multicolumn{5}{c}{RPBE+D3}\\
Limiting potential  & 0.66 & 0.84 & 0.93 & 0.89\\
Overpotential & 0.56 & 0.39 & 0.30 & 0.34\\
\end{tabular}
\end{ruledtabular}
\end{table}
\section{Summary}
\label{sec:summary}
In this work, we have used five different GGA functionals (PBE, PBE+D3, RPBE, RPBE+D3, and BEEF-vdW) to assess the consistency and accuracy of predicted ORR activity of the Fe-$\mathrm{N_{4}}$-C and Co-$\mathrm{N_{4}}$-C catalysts.
It is found that the contribution of dispersion interaction is significant and that RPBE+D3 predicts binding energies, limiting potentials, and the potential determining steps with similar accuracy to BEEF-vdW on both Fe-N$_4$-C and Co-N$_4$-C active sites.
In addition to the pristine Fe-N$_4$-C and Co-N$_4$-C catalysts, we investigate the OH-terminated ones and found that there is a slight lowering of the predicted $\eta_{\mathrm{ORR}}$ value for the case of Fe-$\mathrm{N_{4}}$-C, while no significant change in Co-$\mathrm{N_{4}}$-C.
Our results suggest that further investigation is required to clarify the role of the OH termination, which is supposed to improve the catalytic activity of Co-N$_4$-C active site.
%
Other functionalization groups, different N and C coordinations/configurations, and defects in the vicinity of the active site should also be considered for comprehensive understanding of the single-atom catalysts.
%
We anticipate that our theoretical assessment will be useful in selecting the appropriate XC functional for studying catalytic systems and will serve as a basis for more realistic simulations including the solvent and potential by using the state-of-the-art hybrid DFT and implicit solvation theory \cite{Nishihara2017,Haruyama2018} and for multiscale simulations based on the microkinetic analysis \cite{tripkovic2010oxygen,wang2019self,rebarchik2020noninnocent} toward more accurate description of the catalytic activity of the single-atom catalysts.
\begin{acknowledgments}
This work was partly supported by Grant in Aid for Scientific Research on Innovative Areas "Hydrogenomics" (Grant No. JP18H05519)
%
and “Program for Promoting Researches on the Supercomputer Fugaku” (Fugaku battery \& Fuel Cell Project)
from the Ministry of Education, Culture, Sports, Science, and Technology, Japan (MEXT).
A.F.Z.A. acknowledges the financial support from MEXT.
A part of calculations was performed using the facilities of the Supercomputer Center, the Institute for Solid State Physics, the University of Tokyo, and of the Cybermedia Center, Osaka University.
\end{acknowledgments}



\bibliography{references}

\end{document}